\documentclass[twocolumn]{aastex631}

\begin{document}

\title{What holes in the gas
  distribution of nearly face-on galaxies can tell us about the host disk parameters: the case of the
  NGC 628 South-East superbubble}

\author[0000-0003-2808-3146]{S. Jim\'enez}
\affiliation{Instituto Nacional de Astrof\'isica, \'Optica y Electr\'onica, AP 51, 72000 Puebla, M\'exico.}

\affiliation{Astronomical Institute, Czech Academy of Sciences, Bocni II 1401, CZ-141 31 Prague, Czech Republic.}
\author[0000-0002-3814-5294]{S. Silich}
\affiliation{Instituto Nacional de Astrof\'isica, \'Optica y Electr\'onica, AP 51, 72000 Puebla, M\'exico.}

\author[0000-0002-4677-0516]{Y.D. Mayya}
\affiliation{Instituto Nacional de Astrof\'isica, \'Optica y Electr\'onica, AP 51, 72000 Puebla, M\'exico.}

\author[0000-0001-8216-9800]{J. Zaragoza-Cardiel}
\affiliation{Instituto Nacional de Astrof\'isica, \'Optica y Electr\'onica, AP 51, 72000 Puebla, M\'exico.}
\affiliation{Consejo Nacional de Humanidades, Ciencias y Tecnolog\'ias, Av. Insurgentes Sur 1582, 03940, Ciudad de M\'exico, Mexico.}
\affiliation{Centro de Estudios de F\'isica del Cosmos de Arag\'on (CEFCA), Plaza San Juan, 1, 44001, Teruel, Spain.}



\begin{abstract}
Here we explore the impact of all major factors, such as the non-homogeneous gas distribution, galactic rotation and gravity, on the observational appearance of superbubbles in nearly face-on spiral galaxies. The results of our 3D numerical simulations are confronted to the observed gas column density distribution in the largest South-East superbubble in the late-type spiral galaxy NGC 628. We make use of the star formation history inside the bubble derived from the resolved stellar population seen in the HST images  to obtain its energy and demonstrate that the results of numerical simulations are in good agreement with the observed gas surface density distribution. We also show that the observed gas column density distribution constraints the gaseous disk scale height and the midplane gas density if the energy input rate could be obtained from observations. This implies that observations of large holes in the interstellar gas distribution and their stellar populations have the potential power to solve the midplane gas density - gaseous disk scale-height degeneracy problem in nearly face-on galaxies. The possible role of superbubbles in driving the secondary star formation in galaxies is also briefly discussed. 
\end{abstract}


\section{Introduction}
\label{sec1}

Since the pioneer papers by \cite{Heiles1979, Heiles1984, Heiles1987} and \cite{Brinks1986}, it became clear that shell-like structures and holes are characteristic features of the Interstellar gas distribution in star-forming galaxies.

The large kinetic energies (up to $10^{53}$ erg) of the observed shells, their sizes, shapes and radial distributions led \cite{Bruhweiler1980, Ehlerova1996} to concluded that they are a natural by-product of the stellar feedback on the interstellar medium (ISM) of their host galaxies. These results are in agreement  with \cite{1975ApJ...200L.107C}, who suggested that the gas ejected by massive stars is heated up to large temperatures ($10^{6}$ K - $10^{7}$ K) within the bubble volume due to multiple stellar wind collisions and supernovae (SNe) explosions. This enhances the inner thermal pressure that drives a strong shock into the surrounding ISM and sweeps it up into a dense shell.

Indeed, diffuse X-ray emission has been detected around young star-forming regions in the Large Magellanic Cloud \citep[LMC, e.g][]{1990ApJ...365..510C,2001ApJS..136..119D, 2002AJ....123..255O}. \cite{2011AJ....141...23B} detected more than 1000 HI holes in a sample of nearby galaxies studied within ``The HI Nearby Galaxy Survey'' (THINGS) project. The hole sizes range from $\sim 100$ pc to about 2 kpc while the ages of the embedded stellar populations were estimated to be in the range of (3-150) Myr. 

\cite{McLeod2020} made an important step forward in systematic studies of the stellar feedback on the host galaxy ISM. They presented MUSE integral field unit (IFU) observations of the nearby ($\sim 2$ Mpc) dwarf spiral galaxy NGC 300. These observations, in combination with Hubble Space Telescope (HST) photometry, allowed them to obtain characteristics of individual massive stars in five large HII regions and study the impact of stellar feedback on the ambient ISM in this particular case.  \cite{Nath2020} studied the superbubbles size distribution and showed by means of numerical simulations that the largest superbubbles are likely driven by multiple OB associations and can reach $\sim 1$ kpc in tens of million years. JWST Mid-Infrared Instrument (MIRI) observations now allows us to identify bubbles driven by young stellar clusters by tracing emission of the Polycyclic Aromatic Hydrocarbon (PAHs) molecules associated to shells around the bubbles \citep{Rodriguez2023,2023ApJ...944L..24W}.

Relatively little attention has been paid to the impact of the host galaxy disk rotation and inclination on the study of superbubbles. It is probably because this requires time-consuming 3D hydrodynamical simulations. The effects of the disk rotation were first discussed by \cite{1990A&A...227..175P} who found, by means of 2D calculations, that at later times shells are distorted by the differential galactic rotation and the swept-up mass is concentrated at the opposite tips of the wind-driven shell. 3D simulations by \cite{silich1992} confirmed these findings and also showed that in the case of a plane-stratified interstellar gas distribution the majority of the swept-up mass is concentrated in the bubble belt - a thin zone of the expanding shell next to the midplane of the host galaxy. 

\begin{figure}
\includegraphics[width=\columnwidth]{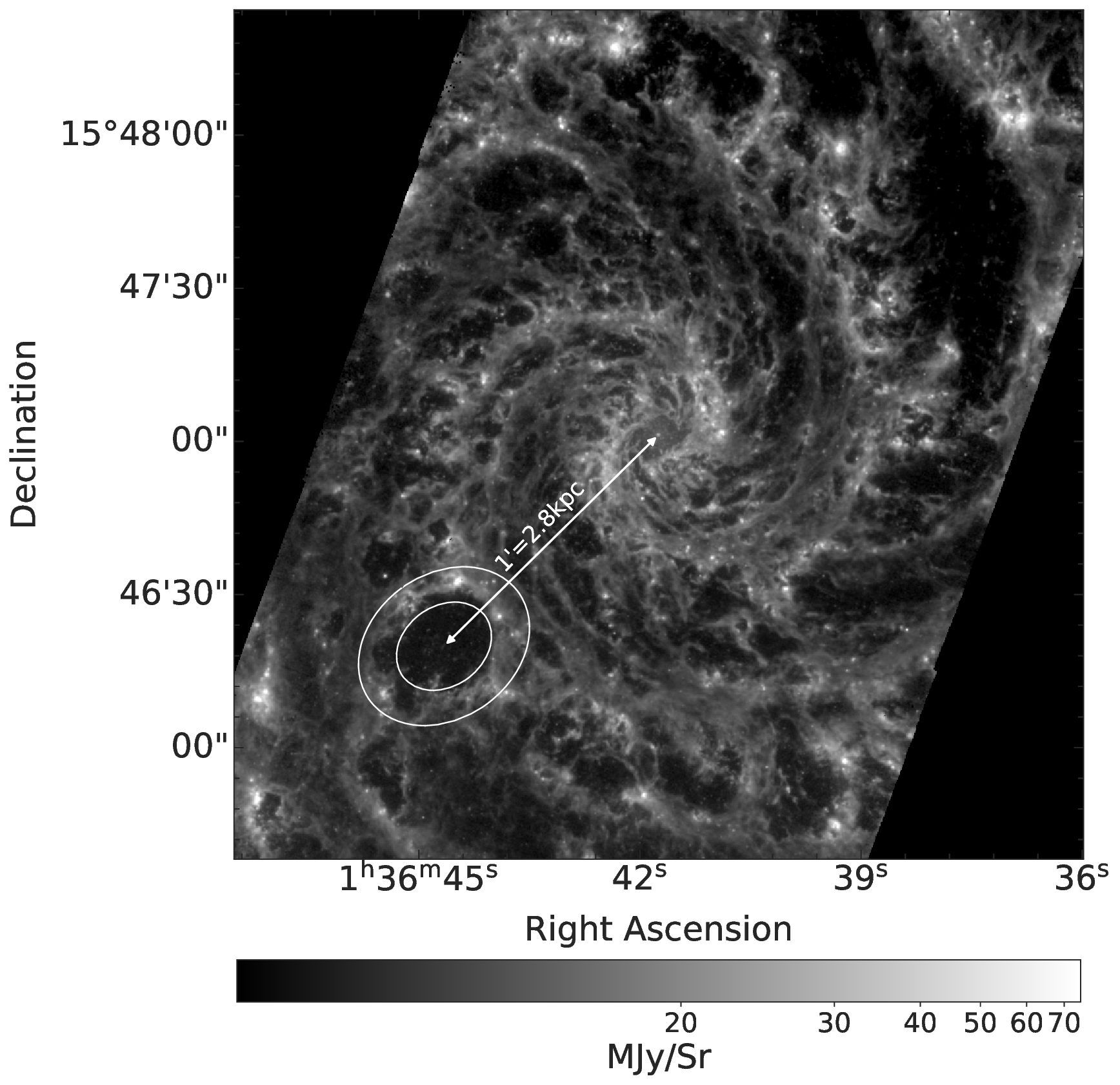}

  \caption{The JWST/MIRI F770W image showing the largest, kpc-size hole in the
gas column density distribution in
   the South-East zone of NGC 628. The inner and outer edges of the
region with an enhanced gas column density around the hole are
displayed by
the white ellipses. The radial vector joining the hole center to the
center of the galaxy is also shown.}
\label{NGC628_bubble}
\end{figure}

NGC 628 is a late-type nearby face-on spiral galaxy at a distance of about 9.8 Mpc \citep{2019ApJ...887...80K, 2021MNRAS.501.3621A}. It was extensively observed in the past \citep[e.g.][]{2011MNRAS.410..313S,2015ApJ...815...93G} and recently within the frame of the PHANGS (Physics at High Angular resolution in Nearby GalaxieS) survey \citep[e.g.][]{2021ApJS..257...43L, 2022A&A...659A.191E, 2022ApJS..258...10L}, THINGS \citep{Walter2008} and the PHANGS - JWST Cycle I treasury program \citep{2023ApJ...944L..17L}. It presents a rich population of bubbles and holes whose radii vary from tens to thousand parsecs (see also \citealt{2020AJ....160...66P}). \cite{2023MNRAS.521.5492M} and \cite{Barnes2023} thoroughly discussed the multi-band properties of the largest, kiloparsec-size, South-East superbubble of NGC 628 (see Fig. \ref{NGC628_bubble}) using multi-band JWST, HST, ALMA and VLT observations.
The shell that surrounds this superbubble is traced in CO, PAHs, HI and H$\alpha$ emission. Molecular gas dominates over the neutral and ionized gas components. The shell total mass and diameter are $\sim 2 \times 10^7$ M$\odot$ and $\sim 1$ kpc, respectively. \cite{2023MNRAS.521.5492M} determined the star formation history inside the bubble and in the surrounding stellar disk and concluded that the mechanical power of the enclosed stellar population exceeds that required to form a spherical shell of such mass and size. In Fig. \ref{NGC628_bubble} we show a JWST/MIRI image of NGC 628 in the F770W filter, which trace the bubble.  

Another issue that requires careful analysis is the observed shell thickness. \cite{2023MNRAS.521.5492M} showed that the observed shell is well resolved in the JWST/MIRI images and that it is rather thick ($\sim 200$ pc) when compared with the bubble size ($\sim 1$ kpc). This seems to be in conflict with the standard wind-driven bubble theory. Indeed, the gas density behind an adiabatic bubble-driven shock is about $4 \times \rho_{ISM}$, where $\rho_{ISM}$ is the gas density of the ambient ISM. Hence, the shell thickness should be about $\Delta R = R / 12$ as the swept-up mass is located within the shell. In more realistic, radiative shock models, the shell should be much denser and thinner. Does this imply that the stellar feedback-driven bubble model is not an appropriate explanation for the observed structure? We demonstrate by means of 3D numerical simulations that this is not the case and that the observed thick column density distribution is a natural by-product of a stratified interstellar gas distribution.

Here we present the results of comprehensive 3D numerical simulations of the evolution of superbubbles in a non homogeneous, disk-like ISM. The simulations include the ISM differential rotation, galactic disk inclination and the star formation history obtained from the analysis of the stellar population detected in the HST/ACS and JWST/NIRCam images. It is shown that the energy input rate determined by the star formation history within the South-East superbubble is consistent with the observed bubble properties. The comparison of the model-predicted projected gas column density distribution with the observed one allowed us to determine the galactic gaseous disk properties: its midplane density and vertical scale. The comparison of the hydrodynamic model predictions with the observed bubble properties thus allows one to resolve the midplane gaseous disk density - disk height scale degeneracy in nearly face-on galaxies.

The paper is organized as follows. The model setup is presented in section \ref{ModSetup}. Specifically, the model adopted for the interstellar gas distribution, the adopted rotation curve and the
gravitational field used in the simulations are discussed in Section \ref{disk_gravity}. In Section \ref{sec3} we make use of the star formation history obtained by \cite{2023MNRAS.521.5492M} to derive the energy input rate responsible for the bubble expansion. In Section \ref{sec4} the hydrodynamic scheme used for simulations is presented.  The impact of the input parameters on the column density distribution is discussed in section \ref{sec5}. Here we also present the model that best fits the observations. The major results and conclusions are summarized in Section \ref{conclu_section}.

\section{Model setup}\label{ModSetup}
The theory of wind-driven bubbles was developed by \cite{1975ApJ...200L.107C, 1977ApJ...218..377W, Tomisaka1986, Tomisaka1988, maclow1988, Koo1992, Suchkov1994} and others \citep[see for a review][]{1988ARA&A..26..145T, bisno1}, who considered the thermalized kinetic energy of individual stellar winds and SNe to be the major driving mechanism responsible for the formation and evolution of interstellar bubbles. The impact of additional physical processes, such as radiation pressure and even cosmic rays on the bubble dynamics and observational appearance has been considered later \citep[see reviews by][]{2019ARA&A..57..227K, 2020AJ....160...78R}.

Here we make use of the thin-shell approximation (section \ref{sec4}) to follow the superbubble evolution and shape transformation in a non-homogeneous galactic interstellar medium (ISM) under the influence of the galactic gravity and the ambient gas rotation (section \ref{disk_gravity}). We also estimate the superbubble driving energy from the star-formation history derived from HST observations (see section \ref{sec3}).

\subsection{The adopted interstellar gas distribution and rotation curve}
\label{disk_gravity}

Hereafter we adopt a Gaussian interstellar gas distribution in the galactic disk and assume that the disk component is surrounded by a low density galactic halo: 
\begin{equation}\label{eq1a}
  \rho_{gas}(x,y,z) =  \rho_{0} \exp \left[-z^{2}/H_{z}^{2} \right]
    +  \rho_{h} ,
\end{equation}
where $\rho_{h}=\mu n_{h}$ is the gas volume density in the galactic halo, $\rho_{mp} = \rho_0 + \rho_{h}$ is the midplane gas density, $H_{z}$ is the gaseous disk scale-height, and $\mu=14/11 m_{H}$ is the mean mass per particle in the neutral gas with 10 hydrogen atoms per each helium atom.

The components of the gravitational field are \citep[e.g][]{1989MNRAS.239..605K, bisno1,2018A&A...619A.101E}:
\begin{equation} \label{eq2a}
  g_{x}=-\frac{V_{rot^2}}{R^2}x
\end{equation}

\begin{equation}\label{eq2b}
  g_{y}=-\frac{V_{rot^2}}{R^2}y
\end{equation}
\begin{equation}\label{eq2c}
  g_{z}=-2\pi G z \left[\frac{\Sigma_{D}}{\sqrt{z^{2} + H_d^{2}}} +
  2 \rho_{h}\right],
\end{equation}
where $x,y$, and $z$ are Cartesian coordinates, $G$ is the gravitational constant, $V_{rot}$ is the rotation velocity, $R^{2}=x^2+y^2$ is the distance along the galactic plane, $H_d$ and $\Sigma_{D}$ are the stellar disc scale-height and surface density.

The parameters $n_{h}$, $n_0$, and $H_{z}$ are not fixed by the available observations. They are determined by the best numerical fit to the observed gas column density distribution.

\subsection{The energy input rate}
\label{sec3}

The star formation history within the South-East superbubble was carefully analysed by \cite{2023MNRAS.521.5492M} and \cite{Barnes2023}. In  order to determine the bubble-driven energy and the mass deposition rate, we approximate the excess of the star formation rate (SFR) within the bubble over that in the disk (Fig. 13 top in \citealt{2023MNRAS.521.5492M}), as a sequence of $N_{tot}$ instantaneous starbursts separated by $\Delta t = t_{max} / N_{tot}$ time intervals, where $t_{max}\sim 50$ Myr is the time passed since the superbubble starburst onset till now (see \citealt{2002AJ....123.2438S}). 

One can obtain then the corresponding mechanical luminosity and mass deposition rate at any time $t = i \Delta t$, where $i \leq N_{tot}$, by adding the mechanical luminosities and mass deposition rates from all previous mini-starbursts:
\begin{eqnarray}
      \label{eq6a}
      & & \hspace{-0.5cm}
      L_b(t) = \sum_{j=0}^{j<i} L_j(t^{*})
       \\[0.2cm]
      & & \hspace{-0.5cm}
      \label{eq6b}
{\dot M}_b(t) = \sum_{j=0}^{j<i}{\dot M}_j(t^{*}) ,     
\end{eqnarray}
where $t^{*} =\left( i-j-1\right)\Delta t$ is the time interval between the evolutionary time $t = i \Delta t$ and one of the previous mini-starbursts that occurred at time $t_j = j \Delta t$. 

The mass ${\dot M}_j(t^{*})$ and energy $L_j(t^{*})$ input rates of the mini-starbursts were obtained from the  stellar population synthesis model STARBURST99 \citep{1999ApJS..123....3L}, which includes stellar winds and supernovae upon the assumption of a solar metallicity starburst and a canonical Kroupa initial mass function with lower and upper cutoffs of $M_{low} = 0.1$M$_{\odot}$ and $M_{up} = 100$M$_{\odot}$, respectively. 

\begin{figure}
\includegraphics[width=\columnwidth]{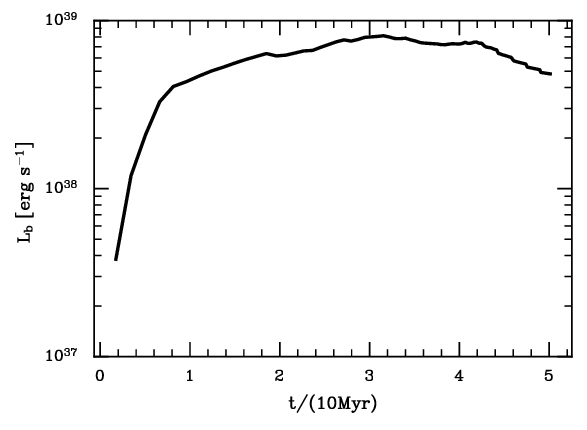}
\caption{The excess of the superbubble  mechanical luminosity over that
in the galactic disk as a function of the superbubble age.}
\label{fig1}
\end{figure}

The mass of each mini-starburst was calculated as:
\begin{equation}\label{eq7}
M_j = \textrm{SFR}(t_j) \Delta t.
\end{equation}
The resulting mechanical luminosity as a function of the bubble age is presented in Fig. \ref{fig1}. The energy input rate first grows fast and then remains almost constant until the SFR becomes negligible at the age of about 50 Myr. This is a cumulative effect of the different age stellar generations detected within the superbubble volume. The total stellar mass formed during 50 Myr is $M_{tot} = \sum_{j=0}^{j=N_{tot}} M_j \sim 10^5$ M$_{\odot}$.

\cite{2023MNRAS.521.5492M} found that recent star formation is not confined to just the inside-bubble zone (see their Fig. 12), instead any region of the disk in the vicinity of the bubble has also been forming stars over the last 50 Myr. Thus the increasing inner bubble volume already contains some stars whose feedback should be considered. Therefore we adopt as the bubble driving energy the sum
\begin{equation}\label{eq8}
L(t) = L_b(t) + L_d(t) \times S_b(t),
\end{equation}
where $L_d(t)$ and $S_b(t)$ are the stellar disk mechanical luminosity \citep[red line in Fig. 12 from][]{2023MNRAS.521.5492M} normalized to the unit surface area and the surface of the bubble cross-section by the galactic plane at a given time t, respectively. The disk component contribution $L_d(t)$ to the bubble mechanical luminosity is calculated in the same manner as $L_b(t)$. However, the second term in equation (\ref{eq8}) also depends on the bubble cross-section and therefore must be calculated at each time step during the simulations.
\subsection{Main equations}
\label{sec4}
All simulations were performed with a 3D code based on the thin shell numerical scheme presented in \cite{1992eoim.conf...65P, silich1992, 2018A&A...619A.101E, 2021MNRAS.505.4669}. There are two major assumptions in this approach. The first one is that all swept-up interstellar gas is accumulated in a thin shell behind the leading shock. The second one is that the inside-bubble thermal pressure $P_{th}$ is uniform. The shell is split into a number of Lagrangian elements and the equations of mass, momentum and energy conservation are solved for each Lagrangian element:
\begin{eqnarray}
      \label{eq9a}
      & & \hspace{-0.5cm}
      \frac{dM_i}{dt}=\rho_{gas}\left(x,y,z\right)
      \left(\mathbf{U_i}-\mathbf{V}\right)\mathbf{n_i} \rm{d} \Sigma_i
       \\[0.2cm]
      & & \hspace{-0.5cm}
      \label{eq9b}
      \frac{d \left(M_i \mathbf{U_i} \right)}{dt}=P_{th} \mathbf{n_i} 
      d \Sigma_i + \mathbf{V}\frac{dM_i}{dt} + M_i\mathbf{g}
       \\[0.2cm]
      & & \hspace{-0.5cm}
       \label{eq9c}
       \frac{d E_{th}}{dt} = L_b\left(t \right)-\sum_{i=1}^{N_{shell}}
       P_{th} \mathbf{U_i}\mathbf{n_i} d \Sigma,       
       \\[0.2cm]
      & & \hspace{-0.5cm}
       \label{eq9d}
       \frac{d \mathbf{r_i}}{dt}=\mathbf{U_i}
\end{eqnarray}
where $\mathbf{r_i}$, $M_i$, and ${\rm d} \Sigma_i$ are the i-th Lagrangian element position vector, mass and the surface area, $\rho_{gas}$ is the ambient gas density, $\mathbf{U_i}$, and $\mathbf{V}$ are the Lagrangian element and the ambient gas velocities in the rest frame, $\mathbf{g}$ is the gravitational acceleration, $\mathbf{n_i}$ is the unit vector that is normal to the i-th Lagrangian element surface,  $\Omega_b$ is the bubble volume, $E_{th}$ is the bubble thermal energy and $\gamma=5/3$ is the ratio of the specific heats. $N_{shell}$ is the total number of Lagrangian elements. All these variables are calculated for each Lagrangian element at each time step. The calculations here presented are performed with $N_{shell} \sim 1600$ Lagrangian elements. 

The projected gas column density distributions were evaluated by making use of each Lagrangian element position $\mathbf{r}_{i}$, orientation (determined by the unit vector that is normal to the Lagrangian element surface), mass $M_{i}$ and surface area $d \Sigma_i$.

The inside bubble thermal pressure is:
\begin{equation}
       \label{eq9e}
P_{th}=\left(\gamma-1\right)\frac{E_{th}}{\Omega_b} .
\end{equation}
One can find a detailed description of the thin shell method in \cite{bisno1}. 
 
\section{Results and discussion}
\label{sec5}
In this section we confront our model with the observed gas column density distribution in the largest superbubble at the South - East of NGC 628 (see Fig. \ref{NGC628_bubble}) whose multi-band  properties were thoroughly discussed by \cite{2023MNRAS.521.5492M} and \cite{Barnes2023}. Following \cite{2023MNRAS.521.5492M}, we adopted a galactocentric radius $R \sim 2.8$ kpc for the superbubble center. At this radius $\Sigma_{D}=30$ M$_{\odot}$ pc$^{-2}$ and the disk rotation velocity $V_{rot}$ is approximately $165$ km s$^{-1}$ \citep{Aniyan2018}. We assume that the bubble was formed 50 Myr ago, when the star formation increased at the present bubble center. The SFR decreased thereafter as described by the star formation history (SFH) reported by \cite{2023MNRAS.521.5492M}. However, in the simulations we also took into account the mechanical luminosity of the older disk component that contributes to the bubble energy balance as each generation of massive stars within the superbubble interior supplies energy during $\sim 40$ Myr, when the least massive ($\sim 8$ M$_{\odot}$) stars explode as supernovae (see section \ref{sec3}). 
\begin{figure*}
\centering

	\includegraphics[width=2.0\columnwidth]{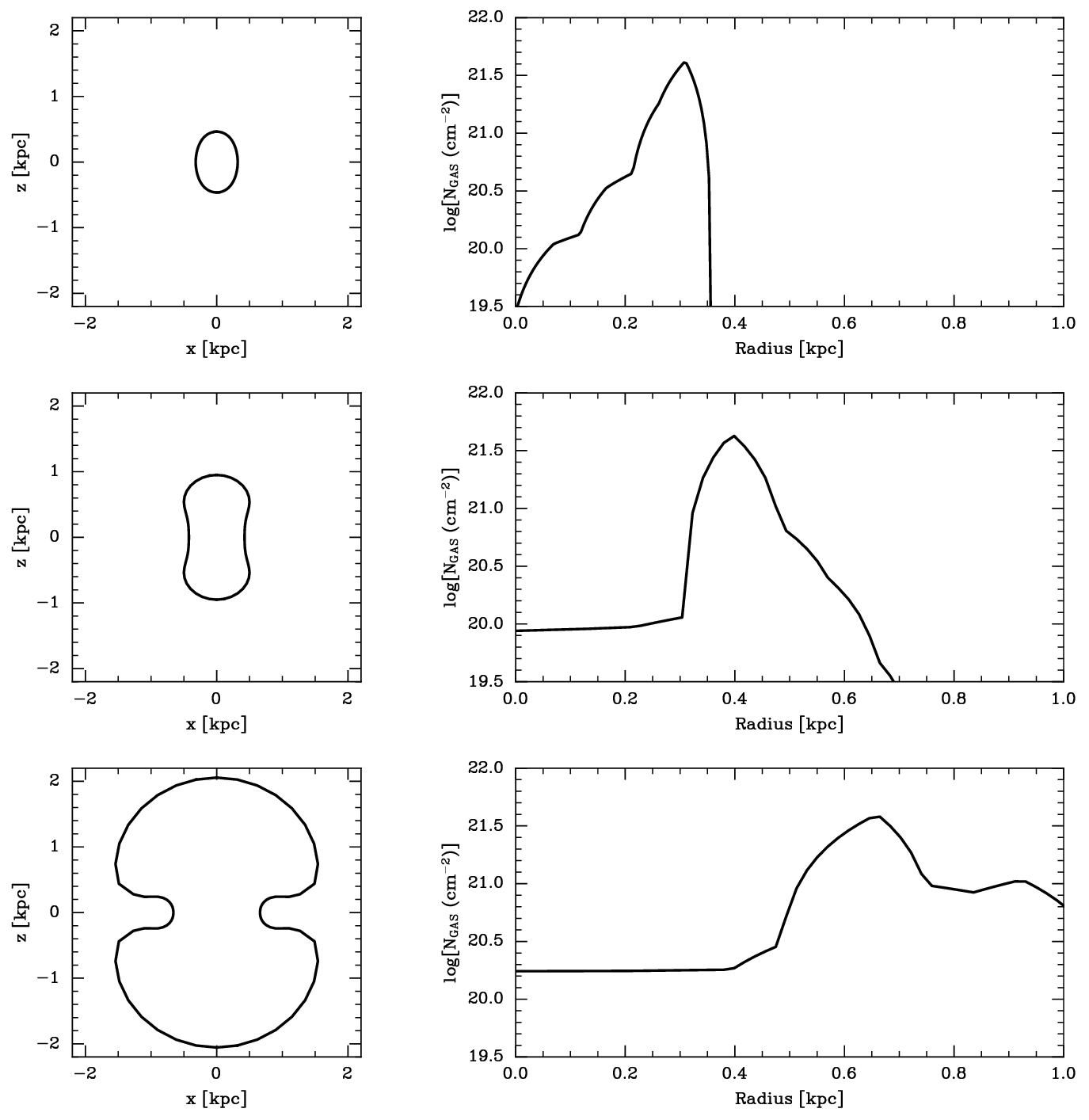}
        \caption{The reference superbubble evolution. The left-hand column
          presents the bubble shape at the age 10 Myr (upper panel),
          15 Myr (middle panel), and 35 Myr (bottom panel). The right-hand column
          displays the model-predicted column density distributions
          at the same bubble ages. The input parameters selected for these
          simulations are $n_{0} = 2$ cm$^{-3}$, $H_{z} = 200$ pc, $i=0^{\circ}$, and the energy input rate derived from the star
          formation history in the NGC 628 South-East superbubble (see section
          \ref{sec3}).}
    \label{fig2}
\end{figure*}

Estimates of the stellar disk scale-height in nearly face-on galaxies are difficult and rather uncertain. We adopt $H_d = 400$ pc as a reference value for the NGC 628 stellar disk scale-height \citep[see][]{Aniyan2018}. However, in section \ref{sec5D} we discuss the impact of the stellar disk scale-height and gravity on the projected gas column density distribution. Another input parameter used in our simulations is the galaxy inclination, which does not affect the hydrodynamical simulations but does affect the simulated column density maps. This is discussed below in section \ref{sec5B}. We adopt for the NGC 628 disk inclination $i = 9^{\circ}$ \citep{1992A&A...253..335K, 2008MNRAS.384L..34D, Aniyan2018}.

The remaining parameters, the gas disk density $n_0$, the halo gas density $n_h$, and the scale-height $H_{z}$, are not determined by observations. We vary them to find the best fit to the observed gas column density distribution. 

\subsection{The reference model}
\label{sec5A}

Fig. \ref{fig2} presents the output from a simulation that does not take into consideration the ambient gas rotation and gravity. The selected halo and disk gas densities and the gaseous disk scale-height are $n_{h} = 5 \times 10^{-2}$cm$^{-3}$, $n_0 = 2$ cm$^{-3}$, and $H_z = 200$ pc, respectively. The energy input rate $L(t)$ was obtained from the star formation history as was explained in section \ref{sec3}.

The left-hand column in Fig. \ref{fig2} shows the superbubble cross-section by the x-z plane at the bubble age of 10 Myr (upper panel), 15 Myr (middle panel), and 35 Myr (bottom panel), respectively. 

One can observe that the bubble shape deviates significantly from a spherical one at the age of 15 Myr. By the time it reaches 35 Myr, the bubble acquires a mushroom-like morphology that remains unchanged thereafter, despite the growth in its size. This strongly affects the column density distribution calculated along line of sights which are normal to the galactic plane (see the right-hand column in Fig. \ref{fig2}). Indeed, the gas column density distribution is getting thicker with the bubble age, while its maximum moves further away from the bubble center.

It is interesting to note that the reference model-predicted column density distribution at the bubble age of 35 Myr looks fairly similar to the emission profiles observed in the direction of the NGC 628 largest bubble \citep[see Fig. 5 in][]{2023MNRAS.521.5492M}.

\subsection{Effects of the galactic disk inclination}
\label{sec5B}
Disk inclination and differential galactic rotation destroy the azimuthal symmetry in the simulated gas surface density maps. Therefore we first calculate the  azimuthally-averaged gas column density distributions \citep[see section 3.2 in][]{2023MNRAS.521.5492M} and then convolve them with a Gaussian Kernel to simulate the impact of the beam characteristics on the model-predicted column density distribution:
\begin{equation}
  N_{con}(R_i) = \frac{\sum_{j} N_{mod}(R_j) \times \exp (-(R_j-R_i)^2) /
    (2 \sigma^{2})}{\sum_{k} \exp (-(R_{k}-R_{i})^{2})/(2 \sigma^{2})} ,
\end{equation} 
where $N_{mod}$ and $N_{con}$ are the model-predicted and the convolved gas column densities. The parameter $\sigma$ is determined by the beam FWHM: $\sigma=\textrm{FWHM}/\sqrt{8 \ln 2}$.  We select $\textrm{FWHM} = 100$ pc as this is approximately the FWHM of the CO observations \citep{2021ApJS..257...43L}.

The convolution slightly reduces the peak column density value and makes the column density distribution broader than that obtained in the hydrodynamical simulations. However, its impact on the peak position is insignificant.

\begin{figure}
\includegraphics[width=\columnwidth]{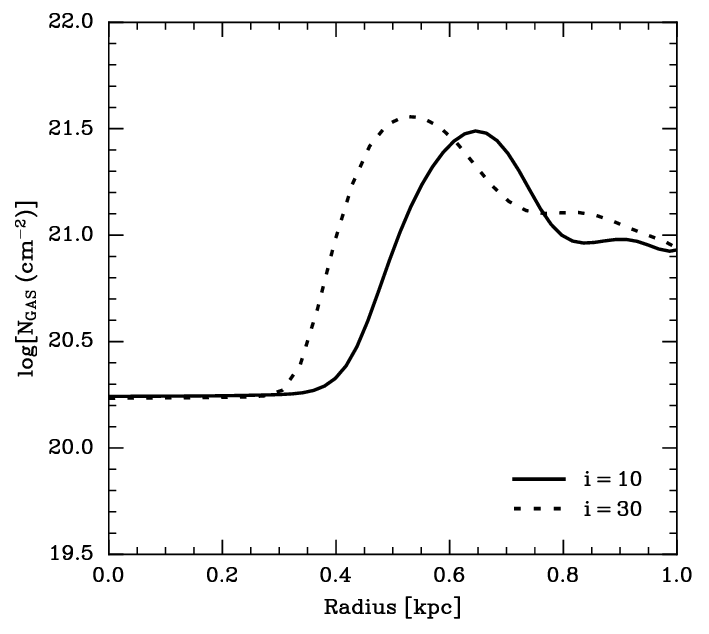}

\caption{The azimuthally-averaged gas column density distribution
after the convolution with a FWHM = 100 pc beam at the age
of 35 Myr. The solid and dashed lines correspond to the disk inclination
angles $i = 10^{\circ}$ and $i = 30^{\circ}$, respectively. All other
input parameters are the same as in the case of the reference model
(see section \ref{sec5A}).}
\label{fig3}
\end{figure}

The impact of the host disk inclination on the column density distribution is shown in Fig. \ref{fig3}, which presents the azimuthally-averaged and convolved surface density profiles obtained upon different assumptions regarding the host galaxy inclination angle. The selected bubble age is 35 Myr. The gas density $\rho_{0}$, the disk scale-height $H_z$ and the energy input rate $L(t)$ used in the simulations are the same as those used in the reference model (see section \ref{sec5A}).

The solid and dashed lines in Fig. \ref{fig3} correspond to the disk inclination angles $i = 10^{\circ}$, $i = 30^{\circ}$, respectively. This plot demonstrates the major effects of the host galaxy inclination. The peak in the column density distribution moves towards the bubble center, the maximum column density slightly increases and the column density distribution becomes wider in models with larger inclination angles.

\begin{figure}
\includegraphics[width=\columnwidth]{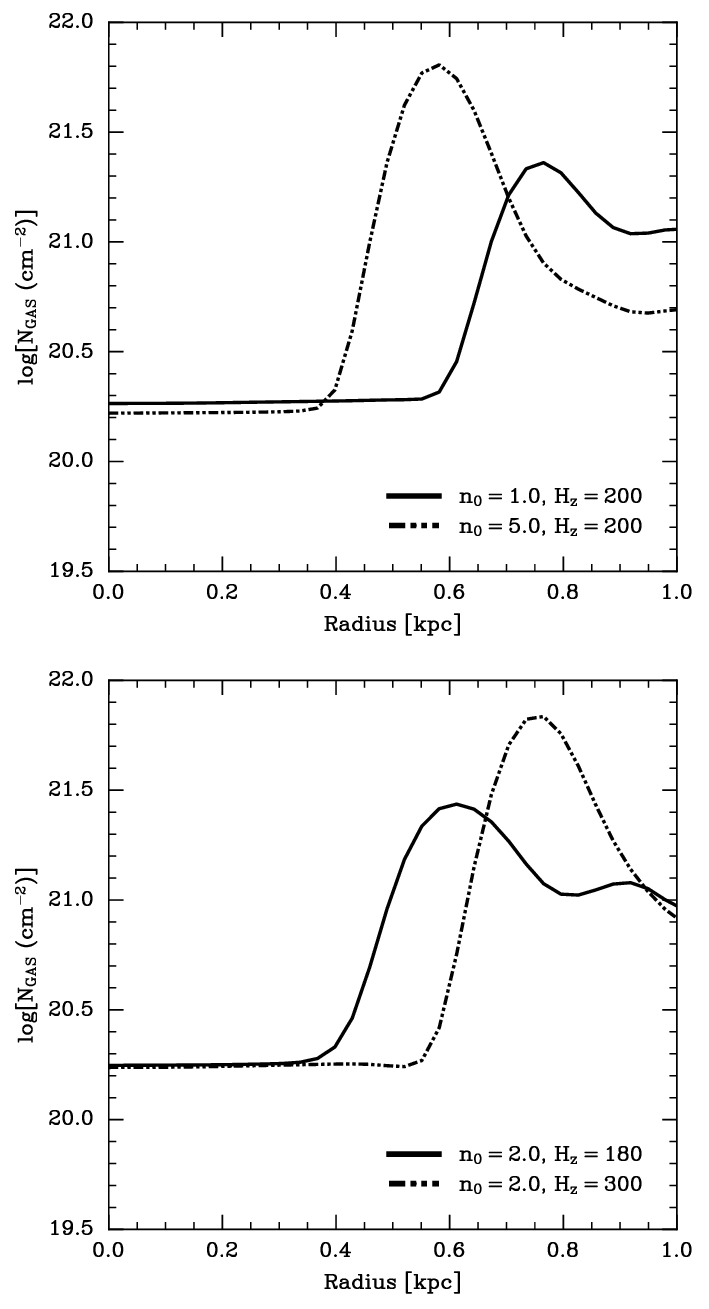}

\caption{The azimuthally-averaged gas column density distribution in
  models with different values of $n_{0}$ and $H_z$ and fixed inclination angle $i=9^{\circ}$ for the host galaxy. The bubble age is 35 Myr. The upper panel displays the
  projected column density distributions in the case when the gaseous disk
  scale-height is fixed to $H_z = 200$ pc, but densities $n_0$ 
  are different: $n_0 = 1$ cm$^{-3}$ (solid line) and $n_0 = 5$ cm$^{-3}$
  (dotted line). The lower panel presents the projected column density
  distributions in the case when the value of the gas density is
  fixed to $n_0 = 2$ cm$^{-3}$, but the disk scale-heights are different:
  $H_z = 180$ pc (solid line) and $H_z = 300$ pc (dotted line), respectively.}
\label{fig4}
\end{figure}

\subsection{Impact of the midplane gas density and gaseous disk scale-height}
\label{sec5C}

We now consider the effects of two input parameters that for face-on galaxies cannot be determined directly from observations because of their degeneracy: these are the midplane gas density $\rho_0$ and the gaseous disk scale-height $H_z$. The impact of the gas midplane density is shown in the upper panel of Fig. \ref{fig4} where we compare the model-predicted column density distributions in cases with $n_0 = 1$ cm$^{-3}$ (solid line) and $n_0 = 5$ cm$^{-3}$ (dotted line) keeping the other input parameters identical to those in section \ref{sec5A}. As one can note, the peak in the column density distribution is larger and moves towards the bubble center in the simulations with a larger midplane gas density. This occurs because the bubble expansion is slower in denser ambient media. 

The bottom panel in this figure shows how the value of the disk scale-height affects the calculated gas column density distribution. The solid line in this panel shows the column density distribution in the case of a smaller scale-height ($H_z = 180$ pc). In the case of the larger scale-height ($H_z = 300$ pc, dotted line) the peak in the column density distribution is located further away from the bubble center and the maximum value of the column density slightly increases. This is because in this case the bubble does not propagate  so rapidly along the z-axis leading to a larger inner bubble pressure and larger cylindrical radii.

These results show that observations of large bubbles in nearly face-on galaxies together with appropriate numerical models have the potential power to solve the midplane gas density - gaseous disk scale-height degeneracy problem. Note that Fig. \ref{fig4} presents simulated column densities convolved with a FWHM = 100 pc beam.

\begin{figure}
\includegraphics[width=\columnwidth]{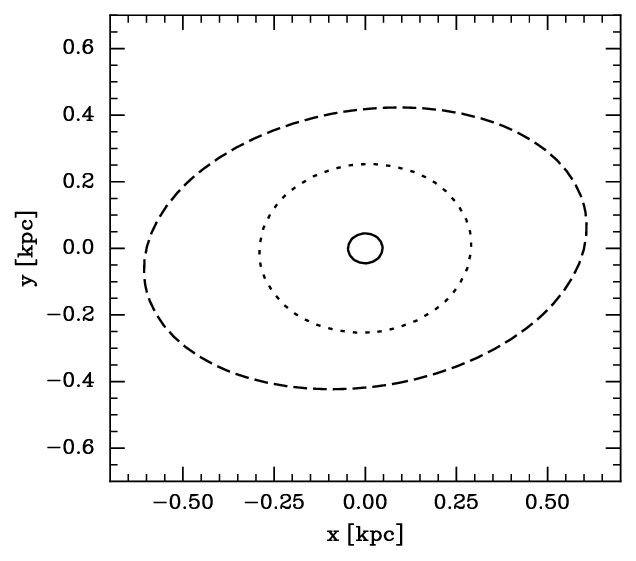}

\caption{ The shell cross-section as seen from above the galactic plane. The adopted input
  parameters are $n_{0}=2$ cm$^{-3}$, $H_{z}=200$ pc, $\Sigma_{D}=30$
  M$_{\odot}$ pc$^{-2}$, $H_{d}=400$ pc, and $V_{rot}=165$ km
  s$^{-1}$. The solid, dotted and dashed lines correspond to the bubble age
  of 10 Myr, 25 Myr, and 50 Myr, respectively.}
\label{fig6a}
\end{figure}
\subsection{Impact of gravity and disk rotation}
\label{sec5D}
Finally, in this section we discuss the impact of gravity and differential disk rotation on the bubble appearance. Two parameters were added to the set of the reference model input parameters: the stellar disk scale height and the gas rotation velocity (see equations \ref{eq2a}-\ref{eq2c} in section \ref{disk_gravity} and equation \ref{eq9b} in section \ref{sec4}). We adopted $H_{d} = 400$ pc, and $V_{rot} = 165$ km s$^{-1}$ \citep[see][]{Aniyan2018}. 

The bubble midplane cross-section evolution is shown in Fig. \ref{fig6a}. Here the solid, dotted and dashed lines display the bubble cross-section shape at the age of 10 Myr, 25 Myr, and 50 Myr, respectively. The other input parameters used in the simulations are: $\Sigma_{D}=30$ M$_{\odot}$ pc$^{-2}$,  $n_{0}=2$ cm$^{-3}$ and $H_{z}=200$ pc. The inclination angle is $i=9^{\circ}$. One can note that the differential galactic rotation distorts the cross-section  shape significantly after about 20 Myr of the bubble expansion. After this time the cross-section obtains an elliptical form and becomes progressively more elongated with the bubble age. In addition, the cross-section semi-major axis rotates with time. 

\begin{figure}
\includegraphics[width=\columnwidth]{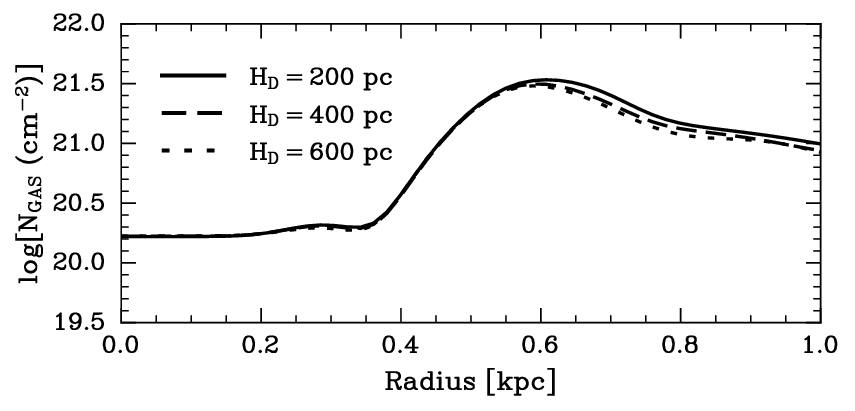}
\caption{The azimuthally-averaged radial column density distribution
  in models with different stellar disk scale-heights at the bubble
  age of 35 Myr. The solid, dashed and dotted lines correspond to
  $H_d=200$ pc, $H_d=400$ pc, and $H_d=600$ pc, respectively. The other
  input parameters are the same as in the model presented in Fig. \ref{fig6a}.
  }
\label{fig6b}
\end{figure}

Fig. \ref{fig6b} presents the projected, azimuthally-averaged and convolved (FWHM = 100 pc) radial column density  at the age of $t=35$ Myr. One can compare the dashed line in Fig. \ref{fig6b} with the right-hand bottom panel in Fig. \ref{fig2} (which presents the results of the simulations with the same input parameters but without gravity), to realize how gravity and galactic rotation affect the results. In the calculations without gravity and rotation presented in Fig. \ref{fig2}, the peak in the column density distribution is located at $\sim 0.65$ kpc, while in the simulations with gravity and rotation it is located closer to the bubble center, at $\sim 0.6$ kpc. In addition, the column density distribution becomes broader in the simulations with gravity and the ambient gas rotation.

We also present in Fig. \ref{fig6b} two models with different stellar disk scale-heights,  $H_{d} = 200$ pc (solid line) and $H_{d} = 600$ pc (dashed line), respectively. One can note that the impact of the stellar disk scale-height on the results is not significant. It slightly changes the bubble expansion velocity along the z-axis and the bubble shape but does not affect the column density distribution significantly (one can note a small difference only at large radii $\ge 0.7$kpc). 

\begin{figure*}
\centering
	\includegraphics[width=2\columnwidth]{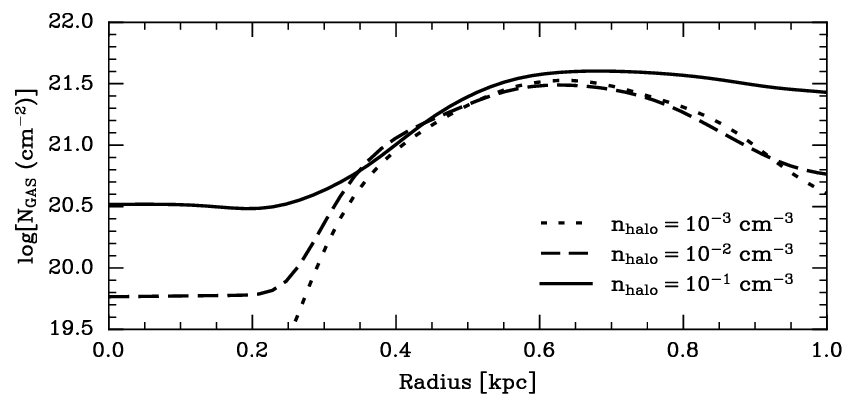}
        \caption{Impact of the halo gas density on the model-predicted
          column density distribution. The solid, dashed and dotted lines
          correspond to $n_{h} = 10^{-1}$cm$^{-3}$,
          $10^{-2}$cm$^{-3}$, and $10^{-3}$cm$^{-3}$, respectively.}

    \label{fig7a}
\end{figure*}
\subsection{Best-fitted model}
\label{sec5E}

We now fix the inclination angle to  $i = 9^{\circ}$, $H_{d}=400$ pc, $\Sigma_{D}=30$ M$_{\odot}$ pc$^{-2}$, $V_{rot}=165$ km s$^{-2}$ \citep{1992A&A...253..335K, 2008MNRAS.384L..34D, Aniyan2018}, and vary the midplane gas density $n_0$ and the scale-height $H_z$ in models with different halo densities $n_{h}$ looking for the best fit to the observed column density distribution in the NGC 628 South-East superbubble. We confront the results of our simulations to the sum of the neutral and molecular gas column densities obtained from the observed, azimuthally averaged HI and CO radial intensity profiles (see Fig. 5 in \citealt{2023MNRAS.521.5492M}), as the contribution of the ionized gas to the total mass is negligible.

To derive column densities from the observed HI intensity we make use of the following relation from \cite{Walter2008}:
\begin{equation}
       \label{eq10a}
       N_{\rm HI} [\textrm{cm}^{-2}] = 1.823 \times 10^{18} I_{\rm HI}.
\end{equation}
The molecular gas column density is:
\begin{equation}
       \label{eq10b}
N_{H_2}[\textrm{cm}^{-2}] = X_{\rm{CO}} I_{\rm{CO}},
\end{equation}
where $X_{\rm{CO}} = 3.3 \times 10^{20} (\textrm{K}\,\textrm{km}\,\textrm{s}^{-1})^{-1}$ and $I_{HI}$ and $I_{CO}$ are the HI and CO intensities expressed as the velocity integrated surface brightness temperatures in units of K\,km\,s$^{-1}$. The value of $X_{CO}$ is that of the Milky Way CO(1-0) \citep{Bolatto2013} taking into account a factor of CO(2-1)/CO(1-0) of 0.61 measured for this galaxy \citep{2021MNRAS.504.3221D}.

Fig. \ref{fig7a} compares the results of the simulations with different halo densities that reasonably reproduce the value of the maximum column density and the column density peak position observed around the NGC 628 South-East superbubble. Larger scale-heights $H_{z}$ are required in simulations with smaller halo gas densities to obtain maximum column densities similar to the observed value and accommodate it near the observed position. Indeed, the required $H_{z}$ increases from about 200 pc in simulations with $n_{h} = 10^{-1}$ cm$^{-3}$ to about 330 pc in simulations with $n_{h} = 10^{-3}$ cm$^{-3}$. However, in simulations with a large halo density ($n_{h} = 10^{-1}$ cm$^{-3}$) the column density distribution looks too flat at large radii. It becomes narrower in simulations with smaller halo gas densities. In these cases the model predicted gas column density also drops significantly in the central ($r \le 0.3$ kpc) zone. In contrast, the value of the halo gas density almost does not affect the required midplane gas density ($n_0 = 2.6$ cm$^{-3}$, 2.4 cm$^{-3}$ and $2.5$ cm$^{-3}$ in models with $n_{h} = 10^{-3}$ cm$^{-3}$, $10^{-2}$ cm$^{-3}$, and $10^{-1}$ cm$^{-3}$, respectively).

\begin{figure*}
\centering
	\includegraphics[width=2\columnwidth]{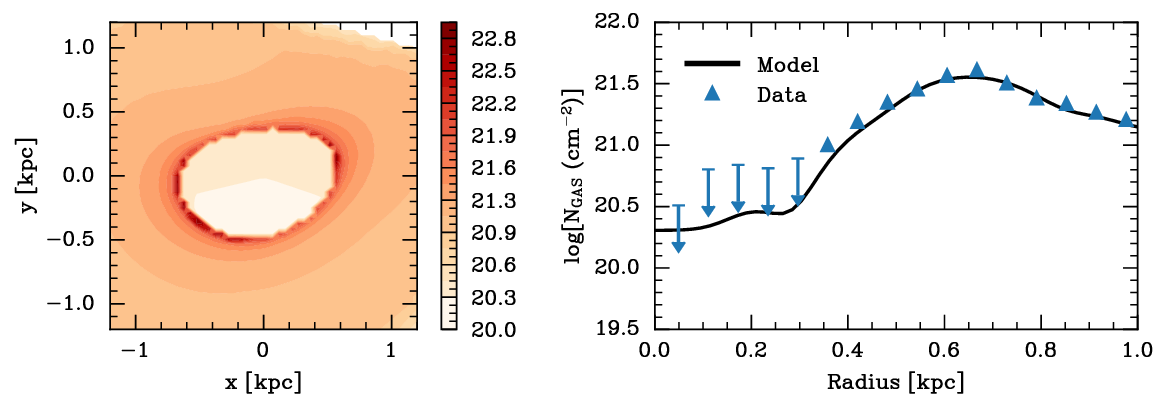}
        \caption{The best-fitted bubble model. The left
          panel presents the projected gas column density map at full numerical resolution in the case
          when the NGC 628 disk inclination angle is $i=9^{\circ}$. The
          corresponding azimuthally-averaged radial gas column density
          distribution after convolving with a beam of $100$ pc is shown in the right panel. The model input parameters
          are $n_0=2.3$ cm$^{-3}$, $n_h=5 \times 10^{-2}$ cm$^{-3}$, $H_{z}=250$ pc,
          $H_{d}=400$ pc, $\Sigma_{D}=30$ M$_{\odot}$ pc$^{-2}$,
          and $V_{rot}=165$ km s$^{-1}$. The results of the
          simulations are presented by the solid line while triangle symbols
          display the sum of the observed neutral and molecular gas column
          densities.}
    \label{fig5}
\end{figure*}
The model that best fits observations is shown in Fig. \ref{fig5}, where the left-hand panel displays the simulated column density map at the bubble age of 50 Myr and the right-hand panel shows the theoretical and the observed azimuthally-averaged radial column density distributions. Here the solid line displays the simulated radial column density distribution after the model results were convolved with a $\textrm{FWHM} = 100$ pc beam, which corresponds to the approximately FWHM=2$^{\prime\prime}$ beam of the CO observations at the distance of NGC 628. The observed column densities are shown by the triangle symbols. The observed H2 + HI column densities in the inner-most parts of the cavity are within the 3-$\sigma$ noise and hence the plotted points correspond to the column density upper limit. 

The model is in excellent agreement with observations at distances $\ge 300$pc from the bubble center, where most of the swept-up gas is located. At smaller radii, the model-predicted column densities also agree with observations as all the model points fall below the upper limits obtained in \cite{2023MNRAS.521.5492M}.

It is interesting to observe that the ellipticity $q=b/a$, where $a$ and $b$ are the semi-major and semi-minor axes of the hole, is $\sim 0.7$ in our calculations, in good agreement with the observed value derived from the star-forming clumps distribution in the shell. Furthermore, the shell diameter along the semi-major axis is $d=2a \sim 1.2$ kpc, also  close to the observed values. 

We also performed numerical simulations with the best-fitted model input parameters upon the assumption of an exponential vertical gas distribution and did not find significant differences with the results presented in Fig. \ref{fig5}.

\begin{figure}
\includegraphics[width=\columnwidth]{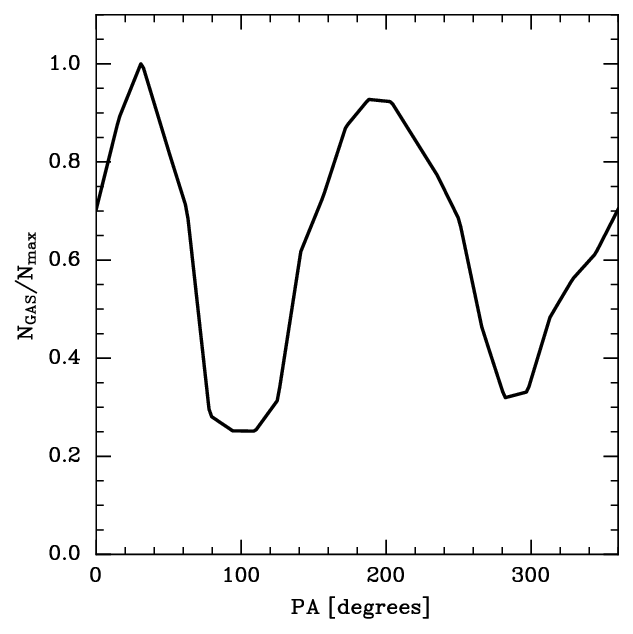}

\caption{ The best-fitted model column density angular distribution.
  The position angle (PA) is measured counter-clockwise from the positive x-axis.}
\label{fig7}
\end{figure}

Another aspect of the differential galactic rotation is shown in Fig. \ref{fig7}. Here we present the  normalized angular column density distribution within the ring 460 pc $\le r \le$ 930 pc, to compare it with the normalized flux azimuthal distribution presented on Fig. 8 by \cite{2023MNRAS.521.5492M}. The position angle (PA) in Fig. \ref{fig7} is measured counter-clockwise from the positive x-axis. The figure clearly demonstrates that the majority of the swept-up matter is accumulated at the opposite tips of the major axis of the oval-shaped bubble-driven shell, within the shell sectors around $ \textrm{PA} \sim 30^{\circ}$ and $\textrm{PA} \sim 200^{\circ}$. It is likely that the interstellar matter accumulation in the opposite tips of the superbubble belt yields in a secondary star formation in these regions. This suggestion is supported by the fact that the positions of the enhancements in the model-predicted  surface density distribution are in remarkably good agreement with peaks in the observed H$\alpha$ emission, which trace the sites of recent star formation \citep[see Fig. 8 in][]{2023MNRAS.521.5492M}.
It is important to highlight that the  comparison of the column density distributions derived from the model and the observed profiles restricts the midplane gas densities and the gaseous disk scale-heights (the best model requires $n_0 \approx 2.3$ cm$^{-3}$, $n_h \approx 5 \times 10^{-2}$ cm$^{-3}$,  and $H_{z} \approx 250$ pc, respectively) and thus may solve the gaseous disk midplane density - scale-height degeneracy.

\subsection{Model uncertainties and simplifications}
It was assumed in our simulations that the bubble expands into a smooth interstellar gas distribution while the NGC 628 galactic disk has a very complex density structure, as evident in the 770W JWST image (see Fig. \ref{NGC628_bubble}). However, it is unlikely that small-scale (in comparison with the superbubble size) inhomogeneities affect our conclusions significantly. It is expected that the bubble-driven shell just overtakes sufficiently smaller bubbles. Certainly, this should result in a more rippled, less smooth shell structure, but it should not affect the shell dynamics significantly. The situation becomes more complicated if the superbubble was not driven initially from a single center, but instead formed via merging of several bubbles comparable in size. We cannot exclude this scenario, but it is difficult to believe that in this case the resulting shell would have an almost perfect elliptical shape (see Fig. \ref{NGC628_bubble}) Collisions with spiral arms can distort the superbubble shape and in addition induce star formation at the sites of collisions. This seems to be happening at the North-West side of the superbubble where the recent star formation is concentrated, but we still do not see a significant shell distortion there, probably because the shell reached the spiral arm only recently. The collisions with spiral arms could be included into the model, but this requires more information regarding the position and density distribution in the spiral arms.

It is worth noting, however, that despite the simplified assumptions, the model predictions align well with observations. This leads us to assert that the gas density stratification, and the known galaxy disk rotation velocity and gravity, are the main factors required for modeling the observational appearance of stellar-feedback-driven bubbles. Furthermore, together with strong constraints on the energy input rate obtained from observations, they allow one to fit the observed properties of the stellar feedback-driven bubbles and obtain the host disk parameters in nearly face-on galaxies.

\section{Summary and Conclusions}\label{conclu_section}

Here we discussed the evolution of superbubbles in galaxies with a disk-like interstellar gas distribution. The impact of different input parameters, such as the host disk inclination, differential galactic rotation and gravity, on the observational bubble appearance was thoroughly discussed. The results of 3D numerical simulations were confronted to the observed properties of the largest, $\sim 1$ kpc in size, South-East superbubble in the nearly face-on spiral galaxy NGC 628.
 
We made use of the star formation history derived from HST/ACS and JWST/NIRCam observations by \cite{2023MNRAS.521.5492M} to obtain the inside-bubble stellar population mechanical power and then performed numerical simulations of the superbubble evolution upon different assumptions regarding the interstellar gas properties.  The simulations showed that the mechanical power of the inside-bubble stellar populations is sufficient to explain the observed, $\sim 1$ kpc, hole size.

We then made use of multiple numerical calculations to find the model that best fits the observed radial column density distribution. For each set of input parameters the results of the simulations were convolved with a FWHM=100 pc beam and confronted to the observed column density distribution. The results show that a certain midplane gas density and a certain gaseous disk scale-height are required to fit the observations. This implies that the comparison of large holes in the interstellar gas distribution and their stellar populations with the results of numerical simulations has the potential power to solve the midplane gas density - gaseous disk scale-height degeneracy problem in nearly face-on galaxies.

Two observational parameters, the position of the peak in the column density distribution and the value of the maximum column density, must be fitted simultaneously to determine the gaseous disk scale-height and the midplane gas density. In the particular case of the NGC 628 South-East superbubble this method led us to conclude that the pre-superbubble midplane disk density and the gaseous disk scale-height are $n_{mp} \approx 2.3$ cm$^{-3}$ and $H_{z} \approx 250$ pc, respectively. The model also predicts a non homogeneous angular column density distribution with the swept-up mass concentrated at the opposite tips of the bubble-driven shell. It is remarkable that the model-predicted enhancements in the azimuthal column density distribution correspond to the sites of the enhanced H$\alpha$ emission which mark the sites where the recent star formation occurred. This favors the hypothesis that superbubbles may trigger a secondary star formation in those zones of the bubble-driven shells where a major fraction of the swept-up interstellar matter is accumulated.

\begin{acknowledgments}
We thank our referee, Dr. Ashley T. Barnes for a detailed report full of valuable comments and important suggestions that helped us to improve the original version of the manuscript. This study was supported by CONAHCYT-M\'exico research grants A1-S-28458 and CB-A1-S-25070. The authors also acknowledge the support provided by the Laboratorio Nacional de Superc\'omputo del Sureste de M\'exico, CONAHCYT member of the network of national laboratories. We also thank Sergio Mart\'inez-Gonz\'alez for helpful suggestions regarding the numerical calculations and Jairo Andres Alzate for discussions during the initial stages of this work and for providing us the star formation history in tabular form.
\end{acknowledgments}

\vspace{5mm}
\facilities{HST, JWST, ALMA, MUSE@VLT, Laboratorio Nacional de Superc\'omputo (LNS), M\'exico.}

\software{numpy \citep{2020Natur.585..357H}, scipy \citep{2020NatMe..17..261V}
          }

\bibliography{jsmz_2023}{}
\bibliographystyle{aasjournal}

\end{document}